\documentclass[rnote]{aa}
\usepackage{txfonts}
\usepackage{epsfig}
\usepackage{subfigure}

\newcommand{\gapprox}{\hbox{\lower .8ex\hbox{$\,\buildrel > \over\sim\,$}}}
\newcommand{\lapprox}{\hbox{\lower .8ex\hbox{$\,\buildrel < \over\sim\,$}}}
\usepackage{times}
\usepackage{natbib}
\usepackage{rotating}
\bibpunct{(}{)}{;}{a}{}{,}

\begin{document}

\title{The evolution of the spatially-resolved metal abundance \\ in galaxy clusters up to $z=1.4$}
\author{S. Ettori\inst{1,2},  A. Baldi\inst{3}, I. Balestra\inst{4}, F. Gastaldello\inst{5, 6}, S. Molendi\inst{5}, P. Tozzi\inst{7}}

\offprints{S. Ettori, \email{stefano.ettori@oabo.inaf.it}}

\institute{
INAF, Osservatorio Astronomico di Bologna, via Ranzani 1, I--40127, Bologna, Italy
\and INFN, Sezione di Bologna, viale Berti Pichat 6/2, I--40127 Bologna, Italy
\and Physics and Astronomy Dept., Michigan State University, East Lansing, MI, 48824 USA
\and INAF, Osservatorio Astronomico di Trieste, via G.B. Tiepolo 11, I--34131, Trieste, Italy
\and INAF-IASF, via Bassini 15, 20133 Milan, Italy
\and Department of Physics and Astronomy, University of California at Irvine, 4129 Frederick Reines Hall, Irvine, CA 92697--4575, USA
\and INAF, Osservatorio Astrofisico di Arcetri, Largo Enrico Fermi 5, I-50125 Firenze, Italy
}

\date{Accepted 1 April 2015}

\titlerunning{The evolution of the metal abundance up to $z=1.4$}
\authorrunning{S. Ettori et al.}

\abstract {We present the combined analysis of the metal content of 83 objects in the redshift range 0.09--1.39, and spatially-resolved in the 3 bins $(0$-$0.15, 0.15$-$0.4, >0.4) \, R_{500}$, as obtained with similar analysis using XMM-Newton data in Leccardi \& Molendi (2008) and Baldi et al. (2012).}
{By combining these two large datasets, we investigate the relations between abundance, temperature, radial position and redshift holding in the Intra-Cluster Medium.}
{We fit functional forms to the combination of the different physical quantities of interest, i.e. ICM metal abundance, radius, and redshift.
We use the pseudo-entropy ratio to separate the Cool-Core (CC) cluster population, where the central gas density tends to be relatively higher, cooler and more metal rich, from the Non-Cool-Core systems.}
{The average, redshift-independent, metal abundance measured in the 3 radial bins decrease moving outwards, with a mean metallicity in the core that is even 3 (two) times higher than the value of 0.16 times the solar abundance in Anders \& Grevesse (1989) estimated at $r>0.4 R_{500}$ in CC (NCC) objects. 
We find that the values of the emission-weighted metallicity are well-fitted by the relation $Z(z) = Z_0 \, (1+z)^{-\gamma}$ at given radius. A significant scatter, intrinsic to the observed distribution and of the order of $0.05-0.15$, is observed below $0.4 R_{500}$. The nominal best-fit value of $\gamma$ is significantly different from zero ($>3 \sigma$) in the inner cluster regions ($\gamma = 1.6 \pm 0.2$) and in CC clusters only.
These results are confirmed also with a bootstrap analysis, which provides a still significant negative evolution in the core of CC systems ($P>99.9$ per cent, when counting the number of random repetitions which provides $\gamma>0$). No redshift-evolution is observed when regions above the core ($r > 0.15 R_{500}$) are considered. 
A reasonable good fit of both the radial and redshift dependence is provided from the functional form
$Z(r,z)=Z_0 \; (1+(r/0.15 R_{500})^2)^{-\beta} \; (1+z)^{-\gamma}$, with $(Z_0, \beta, \gamma) = (0.83 \pm 0.13, 0.55 \pm 0.07, 1.7 \pm 0.6)$ in CC clusters 
and $(0.39 \pm 0.04, 0.37 \pm 0.15, 0.5 \pm 0.5)$ for NCC systems.}
{Our results represent the most extensive study of the spatially-resolved metal distribution in the cluster plasma as function of redshift. 
It defines the limits that numerical and analytic models describing the metal enrichment in the ICM have to meet.}


\maketitle

\section{Introduction}

The hot, thin X-ray emitting plasma in galaxy clusters (i.e. the intra-cluster medium ICM)
is enriched with metals ejected form supernovae (SNe) explosions through subsequent episodes of star
formation and subsequent diffusion through several mechanisms, like, e.g., 
ram-pressure stripping, galactic winds, outflows from Active Galactic Nuclei, 
galaxy-galaxy interactions \citep[e.g.][]{schindler08}.

X-ray observations provide direct measurements of 
the metal abundance in the ICM, as well as their 
radial distribution and variation as a function of time.
These measures represent the `footprint' of cosmic
star formation history and are crucial to trace the effect of SN feedback 
on the ICM at different cosmic epochs \citep[e.g.][]{ettori05,calura07,cora08,fabjan10}.

Several studies have addressed the radial distributions of metals in the ICM
at low redshift \citep[e.g.][]{finoguenov00a,degrandi01,irwin01,
tamura04,baldi07,leccardi08,snowden08}. Few others have constrained the evolution 
of the metal abundance at $z\ga0.3$ (e.g. Balestra et al. 2007; Maughan et al. 2008; Anderson et al. 2009; 
a statistical analysis of the combination of these different datasets is presented in Andreon 2012)
In Baldi et al. (2012), we have presented the XMM-Newton analysis of 39 galaxy clusters 
at 0.4 $<$ z $<$ 1.4, covering a temperature range of 2 $\la kT \la$ 12.8 keV. 
We were able to resolve their abundance in 3 radial bins.

In this work, we combine this dataset with the one presented in \cite{leccardi08}, that includes 44 objects at $z<0.31$, 
with gas temperatures between 2.9 keV and 11.3 keV. The analysis performed on the {\it XMM-Newton} data in the two samples 
is identical and can be statistically combined to probe the ICM abundance as function of radial position and redshift. 

We adopt a cosmological model with $H_0=70$ km/s/Mpc, $\Omega_m=0.3$, and $\Omega_\Lambda=0.7$ throughout. Solar abundance values published in \cite{anders89} are adopted for reference. Confidence intervals are quoted at $1\sigma$ unless otherwise stated.

\section{Data analysis}

\begin{figure*}
\vspace*{-3cm}
\centering
\hbox{
 \includegraphics[width=0.5\textwidth, keepaspectratio]{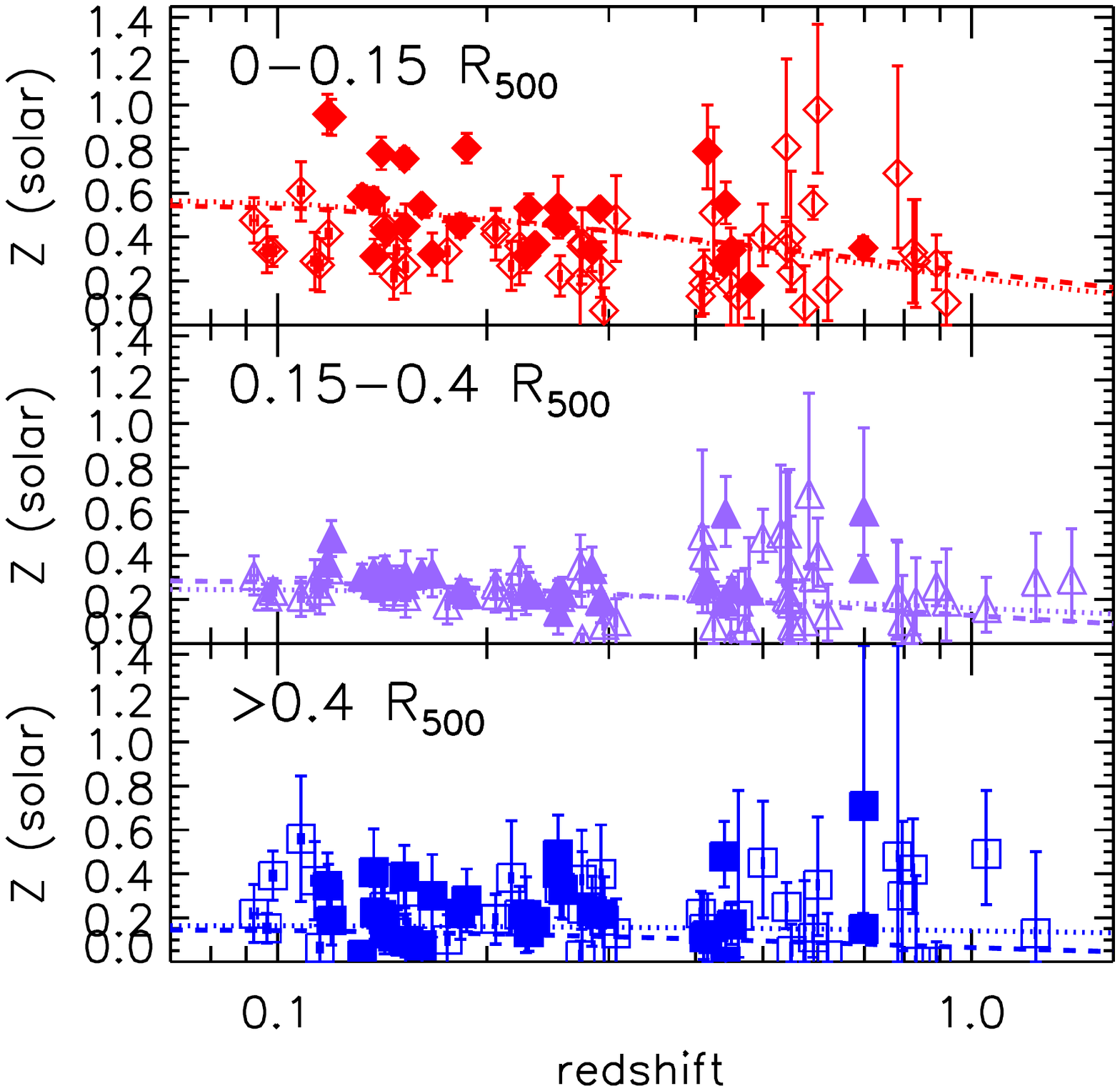}
\includegraphics[width=0.5\textwidth, keepaspectratio]{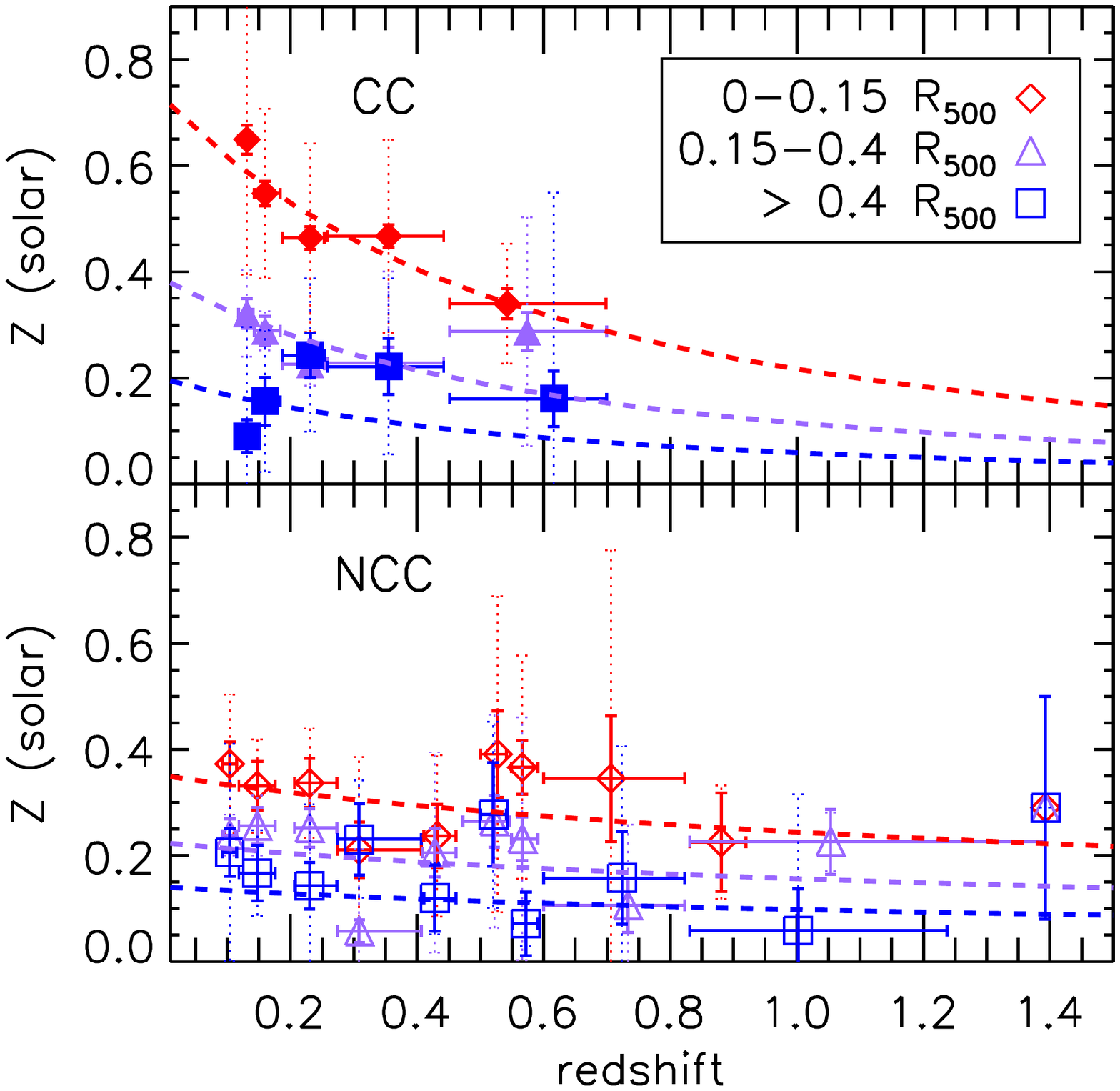}
}
\caption{Abundance distribution as function of redshift as estimated in 3 radial bins ($0$-$0.15 R_{500}$, $0.15$-$0.4 R_{500}$, and $>0.4 R_{500}$). Filled points indicate CC clusters.
(Left) All data are plotted with the best-fits done in the given radial bin (dotted line) and including the radial dependence (dashed line); 
(right) the best-fits in redshift and radius (dashed lines) are shown with representative data points obtained as a weighted mean, 
with the relative error (solid bars) and dispersion (dashed bars).}
\label{z_evol}
\end{figure*}

Details on the X-ray data reduction and analysis are presented in \cite{leccardi08} and \cite{baldi12}, 
where the interested reader can also find a discussion on the systematic effects that could affect our estimates of the ICM metallicity.
We remind that most of these measurements, being for high temperature ($kT > 2$ keV) systems at medium/high redshifts ($z>0.1$), 
are mainly based on the strength of the iron K line (at the rest-frame energies of 6.6-7.0 keV).
The full dataset analysed here is available at {\tt http://pico.bo.astro.it/ $\sim$settori/ abun/xmm.dat}.
We briefly mention here that a simultaneous spectral fit was performed on spectra accumulated from the two MOS detectors over the energy range $0.7-8.0$~keV and using
a Cash statistics applied to the source plus background counts, with a modelled background emission.
The free parameters of the thermal model {\it apec} in {\tt XSPEC} \citep{arnaud96} 
are temperature, abundance, and normalization. Local absorption is fixed to the Galactic neutral
hydrogen column density, as obtained from radio data \citep{kalberla05}, 
and the redshift to the value measured from optical spectroscopy.

Because of the better statistics in lower redshift clusters, the X-ray spectral analysis of most objects in \cite{leccardi08} was performed 
using a finer radial binning with respect to \cite{baldi12}. 
Thus, we had to degrade the spatial resolution of \cite{leccardi08} abundance profiles in order to match its resolution with \cite{baldi12} 
and obtain uniform abundance profiles in three radial bins: $0$-$0.15R_{500}$, $0.15$-$0.4R_{500}$, and $> 0.4R_{500}$, where $R_{500}$ 
is estimated by computing a global gas temperature which does not include the core emission ($<0.15 R_{500}$) and using it in a scaling relation 
(e.g. Vikhlinin et al. 2006; see Baldi et al. 2012 for details).
The spatial degradation was carried out first by interpolating the abundance profile (and its relative error) over the radial points of the observed
surface brightness profile $S(r)$ of each cluster. Then, the surface brightness profile was used as a weighting factor for the interpolated abundance profile, 
together with the errors on the abundance profile itself, so that, in each resulting bin, regions with higher surface brightness 
and smaller uncertainties on the measure of the abundance would have a larger weight in the computation of the average abundance. 
This scheme reproduces an emission--weighted profile of the metal abundance.
Hydrodynamical simulations have shown that an emission--weighted metallicity matches the original value in simulations at better than 5\% 
(e.g.  \cite{rasia08}).
Following this approach, the average abundance $<$$Z$$>$, and the relative error $\sigma_{<Z>}$, in the $R_{min}-R_{max}$ bin was computed as:\\
\begin{equation}
$$
<Z> = \frac{\displaystyle \sum\limits_{i=1}^n \, w_i Z_i}{\displaystyle \sum\limits_{i=1}^n w_i}, 
\hspace*{1cm}
\sigma_{<Z>} = \left( \frac{\displaystyle \sum\limits_{i=1}^n S_i}{\displaystyle \sum\limits_{i=1}^n w_i} \right)^{0.5},
$$
\end{equation}
where $Z_i$ and $\sigma_{Z_i}$ are the abundance profile and the error on the abundance profile, respectively, $w_i = S_i / \sigma_{Z_i}^2$ is the weighting factor,  
and the sum is done over the $n$ radial bins falling in the radial range $R_{min}-R_{max}$ in the abundance profile of each cluster as measured in \cite{leccardi08}.

\begin{table*}
\centering
\caption{Results of the combined analysis on the redshift and radial distribution of the metal abundance.
The columns show: the selected sample; the number of fitted data points; the total $\chi^2$ value; the intrinsic scatter estimated by requiring that the reduced $\chi^2$ is equal to 1; 
the best-fit parameters of the adopted functional form $Z(r,z)=Z_0 \; (1+(r/0.15 R_{500})^2)^{-\beta} \; (1+z)^{-\gamma}$. 
Median, 1st and 3rd quartiles from the distribution of the best-fitting parameters in the bootstrap analysis over a sample of $10^5$ repetitions are indicated in the round parentheses. 
In the last column, labelled ``constant'',  we quote the weighted-mean, its relative error and the dispersion around it in the round parentheses.}
\small{
\begin{tabular}{c@{\hspace{.6em}} c@{\hspace{.5em}} c@{\hspace{.5em}} c@{\hspace{.5em}} c@{\hspace{.5em}} c@{\hspace{.5em}} c@{\hspace{.5em}} c} \hline
 Sample & $N$ & $\chi^2$ & $\sigma$ & $Z_0$ & $\gamma$ & $\beta$ & constant \\
$0-0.15 R_{500}$ & 70 & 267.1 & $0.151 \pm 0.020$ & $0.648 \pm 0.031 (0.647_{-0.073}^{+0.082})$ & $1.60 \pm 0.22 (1.59_{-0.43}^{+0.49})$ & $-$ & $0.451 \pm 0.009 (0.213)$ \\
 only CC & 27 & 140.7 & $0.159 \pm 0.031$ & $0.795 \pm 0.046 (0.780_{-0.095}^{+0.098})$ & $2.19 \pm 0.28 (2.10_{-0.52}^{+0.51})$ & $-$ & $0.493 \pm 0.011 (0.207)$ \\
 only NCC & 43 & 50.6 & $0.053 \pm 0.038$ & $0.360 \pm 0.041 (0.365_{-0.044}^{+0.050})$ & $0.45 \pm 0.40 (0.51_{-0.59}^{+0.71})$ & $-$ & $0.320 \pm 0.019 (0.187)$ \\
 \\
$0.15-0.4 R_{500}$ & 83 & 147.7 & $0.044 \pm 0.012$ & $0.261 \pm 0.022 (0.271_{-0.042}^{+0.035})$ & $0.70 \pm 0.32 (0.66_{-0.49}^{+0.65})$ & $-$ & $0.220 \pm 0.009 (0.134)$ \\
 only CC & 29 & 27.9 & $0.012 \pm 0.025$ & $0.287 \pm 0.032 (0.293_{-0.039}^{+0.072})$ & $0.30 \pm 0.47 (0.33_{-0.58}^{+1.26})$ & $-$ & $0.269 \pm 0.013 (0.120)$ \\
 only NCC & 54 & 96.8 & $0.048 \pm 0.017$ & $0.197 \pm 0.027 (0.218_{-0.074}^{+0.053})$ & $0.37 \pm 0.46 (0.37_{-0.77}^{+0.63})$ & $-$ & $0.178 \pm 0.012 (0.146)$ \\
 \\
$> 0.4 R_{500}$ & 68 & 74.4 & $0.029 \pm 0.028$ & $0.168 \pm 0.028 (0.173_{-0.044}^{+0.068})$ & $0.26 \pm 0.61 (0.33_{-0.82}^{+0.88})$ & $-$ & $0.158 \pm 0.014 (0.170)$ \\
 only CC & 27 & 40.1 & $0.071 \pm 0.038$ & $0.131 \pm 0.031 (0.138_{-0.043}^{+0.106})$ & $-0.88 \pm 0.84 (-0.71_{-1.04}^{+1.41})$ & $-$ & $0.158 \pm 0.019 (0.183)$ \\
 only NCC & 41 & 30.7 & $0.000 \pm 0.008$ & $0.226 \pm 0.053 (0.223_{-0.047}^{+0.056})$ & $1.39 \pm 0.92 (1.31_{-0.94}^{+0.85})$ & $-$ & $0.157 \pm 0.020 (0.164)$ \\
 \\
all & 221 & 506.3 & $0.092 \pm 0.010$ & $0.702 \pm 0.031 (0.699_{-0.079}^{+0.091})$ & $1.33 \pm 0.18 (1.31_{-0.36}^{+0.41})$ & $0.56 \pm 0.03 (0.55_{-0.08}^{+0.09})$ & $0.300 \pm 0.006 (0.181)$ \\
 only CC & 83 & 225.1 & $0.105 \pm 0.015$ & $0.838 \pm 0.046 (0.831_{-0.113}^{+0.127})$ & $1.75 \pm 0.24 (1.72_{-0.53}^{+0.55})$ & $0.55 \pm 0.04 (0.55_{-0.07}^{+0.07})$ & $0.365 \pm 0.008 (0.200)$ \\
 only NCC & 138 & 185.3 & $0.037 \pm 0.013$ & $0.387 \pm 0.038 (0.393_{-0.042}^{+0.043})$ & $0.52 \pm 0.29 (0.53_{-0.42}^{+0.45})$ & $0.39 \pm 0.06 (0.37_{-0.10}^{+0.15})$ & $0.207 \pm 0.009 (0.174)$ \\
\hline \end{tabular}

}
\label{tab:res}
\end{table*}

\section{Results on the spatially-resolved abundance evolution}

This work present our final effort to use archived data to constrain the evolution of the spatially-resolved ICM abundance. 
Our analysis separates between the central regions, where cool cores (if any) appear, and the rest of the cluster atmosphere. 
We require that the spectrum from which the metallicity is measured contains, at least, 300 net counts.
A total of 221 data-points corresponding to the spatially-resolved spectroscopic measurements of the metal abundance distribution in 83 galaxy clusters in the
redshift range 0.092--1.393 is considered.
By estimating the pseudo-entropy ratio $\sigma = (T_0/T_1) \times (EM_0/EM_1)^{-1/3}$, 
where $T_0$ and $T_1$ are the temperatures measured in the $r < 0.15 R_{500}$ region and in the $0.15$-$0.4 R_{500}$ annulus, respectively, 
and $EM_0$ and $EM_1$ are the corresponding emission measures, we define as non-cool-core (NCC) clusters the 54 (out of 83) objects that have $\sigma \ge 0.6$ 
(see e.g.  \cite{leccardi10} and \cite{baldi12b}).
The data points are plotted as function as radius and redshift, separating between CC and NCC systems, in  Fig.~\ref{z_evol}.
All the results are presented in Table~\ref{tab:res}.

In the $r<0.15R_{500}$ radial bin, we have 70 data points over the entire redshift range.
The distribution of the metal abundance correlates significantly with the redshift (Spearman's rank $\rho_z=-0.25$, corresponding to a significance of $2.1 \sigma$ versus the null-hypothesis of no correlation; see Table~\ref{tab:rho}), with the temperature ($\rho_T=-0.3$, $2.5 \sigma$) and with the pseudo-entropy ratio ($\rho_{\sigma}=-0.4$, $3.3 \sigma$).
These correlations point towards a significant effect induced from the cooling activities taking place in these regions, where the ICM at high density radiates more efficiently, lowering its global temperature and increasing its metal budget (see e.g. \cite{leccardi08}). 
Fitting the data points in this bin with a power-law in the form $Z\propto (1+z)^{-\gamma}$, we obtain
$\gamma=1.60 \pm 0.22$, different from zero at high significance ($> 6 \sigma$).
When a bootstrap analysis is done, i.e. the fit is repeated $10^5$ times after a random sampling with replacement and the median, the 1st and 3rd quartiles are considered to estimate location and dispersion of the distribution of the best-fitting parameters, 
the central values are recovered but the relative errors are increased 
by about a factor of 2, mostly because a more proper sampling of the intrinsic scatter is performed with the bootstrap analysis. 
The negative evolution ($\gamma > 0$) is still detected in the $>$99.9 per cent of the replica. 
When we consider CC and NCC systems separately, the normalisation $Z_0$ is different by a factor of 2.2 in favour of CC systems, where a negative evolution is still detected at very high significance.
Instead, no evolution is measured ($\gamma = 0.5 \pm 0.4$) in NCC objects, although the case for negative evolution is still observed in about 80\% of the bootstrapped repetitions.

Eighty-three data points are located in the radial bin $0.15 R_{500} <r< 0.4 R_{500}$, where a $2.8 \sigma$ significant correlation between pseudo-entropy ratio 
and redshift is detected, suggesting an increase of $\sigma$ (i.e larger incidence of NCC systems) with the redshift.
A mild anti-correlation between $Z$ and redshift is also present in this bin (significance of $2.2 \sigma$), 
We measure $\gamma=0.70 \pm 0.32$ ($1 \sigma$ range of $0.16-1.31$ after bootstrap analysis).
The case for negative evolution is still observed in about 91\% of the bootstrapped samples.
Similar values are measured in the population of NCC clusters, which dominates the sample in this radial bin, with a normalisation (and a mean metallicity value) that 
is lower by about 30 per cent than the corresponding value measured for CC systems.

Finally, we do not find evidence of negative redshift evolution of the metallicity also in the last radial bin considered in our analysis ($r>0.4R_{500}$), 
where we constrain a value for $\gamma$ of $0.26 \pm 0.61$ ($\gamma>0$ with a probability $P\approx $ 65\% after bootstrap analysis) with 68 data points.
Similar results are obtained in the CC and NCC populations, where an average value of $Z=0.16 Z_{\odot}$ is estimated.

Being the sample highly heterogeneous and not representative in any sense of the cluster population, it 
is not expected to weight properly for the relative incidence of these cool cores clusters.
Moving at higher redshift, for instance, lower metal abundance associated to the inner regions suggest that 
less pronounced cool cores are resolved in our sample (see e.g. \cite{santos10,samuele11})  

\begin{table}
\centering
\caption{The Spearman's rank correlation coefficient 
(and the significance in terms of the standard deviations by which the sum-squared difference of ranks deviates from its null-hypothesis expected value)
between metallicity and redshift ($\rho_z$), metallicity and temperature ($\rho_T$), metallicity and pseudo-entropy ratio ($\rho_{\sigma}$), and pseudo-entropy ratio and redshift ($\rho_{\sigma, z}$), respectively. The pseudo-entropy ratio is larger in NCC systems.}
 \small{
\begin{tabular}{c@{\hspace{.6em}} c@{\hspace{.5em}} c@{\hspace{.5em}} c@{\hspace{.5em}} c} \hline
 Sample & $\rho_z$ & $\rho_T$ & $\rho_{\sigma}$ & $\rho_{\sigma, z}$ \\
$0-0.15 R_{500}$ & -0.25 (2.07) & -0.31 (2.54) & -0.39 (3.27) & 0.18 (-1.51) \\ 
$0.15-0.4 R_{500}$ & -0.24 (2.18) & -0.13 (1.14) & -0.13 (1.18) & 0.31 (-2.80) \\ 
$> 0.4 R_{500}$ & -0.20 (1.63) & -0.09 (0.75) & -0.10 (0.84) & 0.23 (-1.85) \\ 
all & -0.20 (3.04) & -0.09 (1.27) & -0.20 (2.93) & 0.25 (-3.66) \\ 
\hline \end{tabular}

}
\label{tab:rho}
\end{table}

\section{Summary and Conclusions}

We present the statistical analysis of the spatially resolved spectroscopic results of the {\it XMM-Newton} observations of 83 clusters of galaxies in the redshift range $0.09<z<1.39$ and with a gas temperature between $2$~keV and $12.8$~keV. In a consistent and similar way (see details in \cite{leccardi08} and \cite{baldi12}),
the cluster emission has been resolved for all the objects in the dataset in three different regions: the core region (corresponding
to $0<r<0.15 R_{500}$), the region immediately surrounding the core ($0.15R_{500}<r<0.4R_{500}$), and the outskirts of the cluster ($r>0.4R_{500}$).
Using the pseudo-entropy ratio $\sigma$, we also distinguish between Cool-Core and Non-Cool-Core objects, with CC clusters that represent $\sim$35 per cent of the whole sample.
Our results are summarised in Tables~\ref{tab:res} and \ref{tab:rho}.

We find strong (significance at $3 \sigma$ or higher) correlations between (i) metallicity and pseudo-entropy ratio (in the inner bin) and, in the entire sample, (ii) metallicity and redshift, and (iii) pseudo-entropy ratio and redshift.

By fitting a constant value (under the assumption of no-evolution) to the metal abundance measured in the 3 radial bins, we confirm a neat decrease moving outwards, with the inner bin being even 3 (two) times more metal-rich than the outermost bin in CC (NCC) systems ($0.49/0.32$, $0.27/0.18$ and $0.16/0.16$ for CC/NCC systems in the 3 radial bins, respectively). 

We constrain the parameters of the functional form $Z(r,z)=Z_0 \; (1+(r/0.15 R_{500})^2)^{-\beta} \; (1+z)^{-\gamma}$. 
The best-fitting values are $Z_0 = 0.70 \pm 0.03$ ($0.62-0.79$, 1st--3rd quartile from the bootstrap analysis), $\beta = 0.56 \pm 0.03$ $(0.48-0.65)$ and $\gamma = 1.3 \pm 0.2$ $(0.95-1.72)$.
CC and NCC systems behave differently, with $(Z_0, \beta, \gamma)$ equals to $(0.83 \pm 0.13, 0.55 \pm 0.07, 1.7 \pm 0.6)$ and $(0.39 \pm 0.04, 0.37 \pm 0.15, 0.5 \pm 0.5)$, respectively 
(error bars refer here to the bootstrap analysis).

At a given radius, we find a statistically significant negative evolution in the distribution of the metal abundance as function of redshift in the inner radial bin ($> 3.5 \sigma$; the probability from the bootstrap analysis to measure $\gamma>0$ is larger than 99.9 per cent). 
We show that this level of evolution with the cosmic time is entirely consistent with the same evolution observed in the CC-only population.
On average, we measure a mean metallicity in the local systems that is about 3--4.5 (1.4 in NCC clusters) times the value measured at $z \approx 1$, which is consistent with the constraints we publish in \cite{baldi12} and, using (mostly) {\it Chandra} data, in \cite{balestra07}.
At the same redshift, the metallicity decreases radially, reaching at $\sim 0.5 R_{500}$ about 37 (in CC) and 50 (in NCC) per cent of the value estimated at $0.15 R_{500}$.
By requiring a reduced $\chi^2$ of 1 in the performed fit, we estimate an intrinsic scatter in $Z$ of about $0.15, 0.05$ and $0.05$ in the 3 radial bins, respectively. 
In the inner bin, the scatter of $\sim 0.15$ is measured also in the CC systems only, whereas NCC clusters show a value of $0.05$ consistently in the inner 2 radial bins.

CC clusters show evidence of negative evolution just in the inner radial bin ($<0.15 R_{500}$), where the accumulation of the plasma during the cluster formation makes the ICM denser and, as consequence of radiative processes, cooler and more metal-rich.
Most of the above-mentioned correlations (i.e. $\rho_z$, $\rho_T$ and $\rho_{\sigma}$ in Tab.~\ref{tab:rho}) can be then interpreted as due to the strong influence of this centrally-located cool gas.
NCC systems do not show any statistically significant deviation from $\gamma = 0$ at any radial bin. 
Only in the outer bin ($r >0.4 R_{500}$), both CC and NCC systems present similar average metallicity, and an intrinsic scatter almost consistent with zero. 
We conclude that, for what concerns the metal distribution, this cluster volume is the most homogeneous in both CC and NCC systems, and might be considered the most representative of the average cluster environment.

We stress that the constraints presented in this work represent the most robust ones reachable with archived {\it XMM-Newton} and {\it Chandra} data on the distribution of the metal abundance as function of radius and redshift in the ICM.
Larger samples of high redshift X-ray clusters, together with deeper {\it Chandra} and {\it XMM-Newton} observations even of the clusters already present in the archives, would be crucial to put tighter constraints on the evolution of the ICM metal abundance and to provide a robust modelling of the physical processes responsible for the enrichment of the cluster plasma during its assembly over cosmic time, from the accumulation of the iron mass in the CC cluster cores (e.g. De Grandi et al. 2004, B\"ohringer et al. 2004) to the history of the processes responsible for the metal release into the ICM through ejections from supernovae type Ia and II (e.g. Ettori 2005) and through galaxy transformation in the cluster volume (e.g. Calura et al. 2007), and their modelling in joint semi-analytic/hydrodynamical numerical simulations (e.g. Cora et al. 2008, Fabjan et al. 2010).

\begin{acknowledgements}
We thank Sabrina De Grandi and Mauro Sereno for discussion.
We acknowledge financial contribution from contracts ASI-INAF I/009/10/0 and 
Prin-INAF 2012 on ``A unique dataset to address the most compelling open questions about X-Ray Galaxy Clusters''.
We thank the anonymous referee for the useful comments that helped to improve the presentation of the results in this paper.
\end{acknowledgements}

\bibliographystyle{aa}
\bibliography{Zevol}

\end{document}